\documentstyle[aps,prb,preprint]{revtex}
\newcommand{\beq}{\begin{equation}}
\newcommand{\eeq}{\end{equation}}
\newcommand{\beqar}{\begin{eqnarray}}
\newcommand{\eeqar}{\end{eqnarray}}
\newcommand{\beqars}{\begin{eqnarray*}}
\newcommand{\eeqars}{\end{eqnarray*}}
\newcommand{\req}[1]{(\ref{#1})}

\draft
\title{ Memory and Self Organization }
\author{Matteo Marsili, Guido Caldarelli \\
{\em International School for Advanced Studies (SISSA)\\
V.Beirut 2-4, 34014 Trieste, Italy}}
\date{ }
\begin{document}
\maketitle

\begin{abstract}
The main result of this letter is that SOC naturally arises as a
result of memory effects. 

We show that memory effects provide the mechanism for self
organization. A general procedure to investigate this issue in
models that display self organized critical behaviour is proposed
and applied to some example. The simplest class of models
exhibiting self organized criticality through this mechanism is
introduced and discussed in some detail. 
\end{abstract}


\narrowtext
A large amount of efforts have been recently devoted to uncover
the mechanism underlying the tendency of large statistical
systems to self organize into a critical
state\cite{SOC,kardar,LPSOC,BSRG,FT}. This issue has a great relevance
since self organized criticality (SOC) manifests itself in a large
variety of phenomena ranging from earthquakes\cite{earthq} 
to magnetic systems\cite{aging,bark},
from interface growth\cite{snepp,interf}  to biological
evolution\cite{BS}. 

Much interest has focused on recently proposed models 
that involve quenched disorder\cite{BS,snepp} and whose 
dynamics leads spontaneously to a SOC state. 
The occurrence of critical aspects in connection with a dynamics
in a random environment is not a peculiarity of these
models: invasion percolation\cite{IP} (IP) is known to reproduce
the critical clusters of standard percolation right at the
percolation threshold; non trivial space--time correlations also
appear in spin glasses dynamics\cite{aging}, charge
density waves\cite{cdw} and in zero temperature
dynamics of magnetic systems\cite{bark,RFIM} in quenched disorder. 
Criticality is related to the presence of memory in these
systems. By this we mean that 
the dynamic of local variables is sensible to a
long period of the past history of the process. 

This letter we inquire on the relation between dynamics in
quenched disorder, memory effects and SOC. 
First we discuss how memory arises in models that evolve in
a random environment. Then we show that a SOC behavior does not
necessarily imply the presence of memory. 
We suggest that instead the converse is true: {\em Memory, i.e.
the dependence of local dynamics on the whole history of the
process, implies a self organized critical behavior}. 
This issue is analyzed with the introduction of a
model that contains the effects of memory explicitly. 
In this model there is no reference to quenched random variables. 
For a particular value of the exponent that tunes the effects of
the past history on the evolution, we reproduce the result of the
corresponding dynamics in quenched disorder. The occurrence of
SOC behaviour for a whole range of this exponent suggests
that criticality is not a peculiarity of dynamics in quenched 
disorder but rather it arises as a result of memory effects.

The simplest model of dynamics in quenched disorder 
is perhaps the Bak Sneppen\cite{BS} model (BSM) originally devised to
model biological evolution: assign a uniformly
distributed random variable (RV) $\eta_i$ on each site $i$ of a
$d$--dimensional lattice. At each time step select the smallest
RV and replace it and the RV's on the neighboring sites with
newly extracted uniform RV's. The system self
organizes to a ``critical'' steady state in which almost
all RV's are above a certain threshold value $p_c$.
This state is characterized by long
range correlations, both in space and in time, that have been
studied by different techniques\cite{BSRG,FT,BS}.

Imagine to assign to each site of the lattice a counter variable
$k_i$. At time $t$ the variable $k_i$ is set to zero if the
variable $\eta_i$ is updated and it is increased by one
otherwise. In this way $k_i(t)$ is the time elapsed since the
last update on site $i$. 

In a system that evolves probing a random environment, as the BSM 
and invasion percolation where the extreme statistics of a
random field is selected at each time, it is natural to think
that the evolution will take place more often on recently updated
regions than in older ones. This is because a site whose RV has been 
checked a large $k_i\gg 1$ number of times in the search for the 
minimum RV will probably have a large RV. It still has a probability
of being the smallest in the future but this probability gets smaller
and smaller as time goes on. This implies that the probability
that a site with a counter equal to $k$ is selected decreases with $k$.

It is possible\cite{RTS}, keeping track of the evolution of the
statistics of the RV on each site, to pursue this argument 
further and to evaluate the probability of
each selection event, that is the probability that a random
variable $\eta_i$, that has a counter $k_i(t)$, is the smallest one. 
This in principle provides
a stochastic formulation of the process, that is deterministic for
each realization of the randomness. We will not enter the
details of this (the interested reader is referred to\cite{RTS})
but just note that the probability that site $i$ is the smallest
can be labeled by $k_i$ and evaluated for invasion percolation, 
under approximations of mean field type, with the result\cite{RTS}
\beq
{\rm Prob}\{\eta_i=\min[\eta_j;\,\forall j]\}=\mu_{k_i,t}\sim 
k_i^{-\alpha}
\eeq
for $t\gg k_i\gg 1$ with $\alpha =2$. It is worth to stress here that
$\mu_{k,t}$ is {\em not} a function of $k$ alone in models like
invasion percolation. In a single realization it actually depends on
finer details of the past history. However, on average, it displays a
fairly stable power law dependence on $k$.

The distribution $\mu_{k,t}$ is a directly accessible
quantity in a computer  simulations. Indeed the fraction of
times selection occur on  a site with $k_i=k$ will be
$n_{k,t}\mu_{k,t}$, where $n_{k,t}$ is the number of sites
with counter $k_i(t)=k$ at time $t$.

We will be concerned mainly with the stationary state of 
a system of linear size $L$ with periodic boundary condition. 
In the steady state the above distributions attain a constant value
$\mu_k(L)$ and $n_k(L)$. 
These satisfy the normalization conditions
$\sum_k \mu_k(L) n_k(L)=1$ and $\sum_k n_k(L)=L^d$ for a
system of linear size $L$. 

A measure of the effect of memory is given by the first moment of
the distribution $n_k(L)$ 
\beq
T_t(L)=\frac{1}{L^{d}}\sum_{k=0}^\infty k n_k(L)\sim L^{d(1+\zeta)}.
\label{TL}
\eeq
The exponent $\zeta$ is a measure of the presence of
memory effects. On the average the local dynamic of the variable
on site $i$ is sensible to a period of the past history of the
process. If the length of this period, measured in units of $L^d$
individual events, increases with $L$, i.e. if $\zeta>0$, the state of
the infinite system will depend on the whole history. If
$\zeta=0$ we can say that no memory effect is present.

\begin{figure}
\vspace{8cm}
\includegraphics{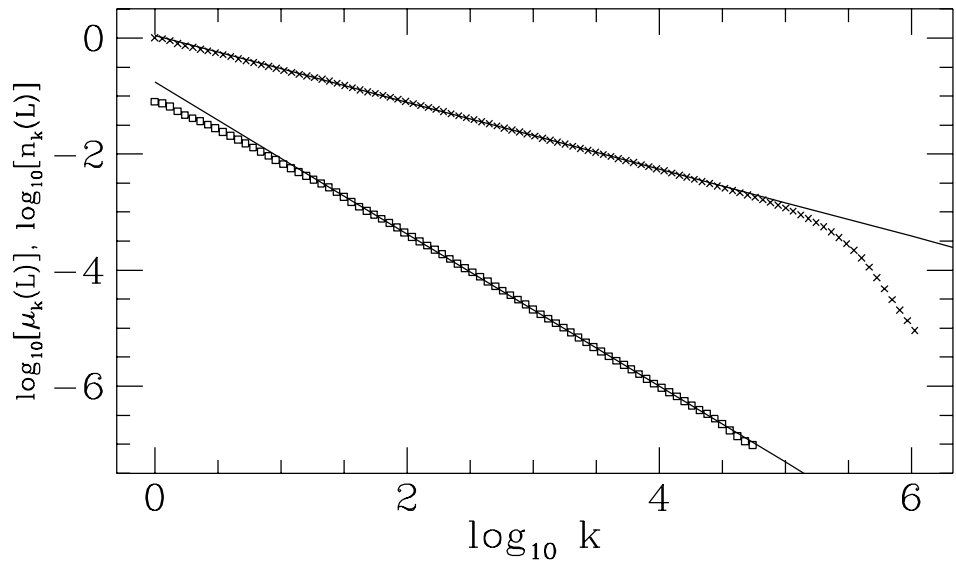}
\caption{The distributions $n_k(L)$ and $\mu_k(L)$ for 
the BSM.}
\label{fig1}
\end{figure}

Counter variables can be introduced in any system. Consider e.g.
the Metropolis dynamics\cite{metro} of the Ising model. A variable is 
selected on average once every $L^d$ attempts. With a probability 
that does not depend on $L$ the move is accepted and the spin 
flipped. Thus we expect $\zeta=0$ for this model and in general
for equilibrium dynamics. 
Consider next the prototype model of SOC, the sandpile 
model\cite{SOC}: sand is added on randomly chosen sites. A 
site cannot store more than $2d-1$ grains and it ``topples''
when it receives the $2d^{\rm th}$ one, i.e. it distributes 
one grain to each of its neighbor sites causing eventually
``toppling'' on these sites as a result. After a toppling a
site is empty. Before it will topple again it needs to store
enough sand. Thus the probability that it will 
topple after $k$ toppling events grows with $k$ and on the 
average it will topple once every $L^d$ toppling events. 
Then $\zeta=0$ also in this case. This result is consistent with
the abelian nature of this model\cite{abel}.
The BSM has instead a non--abelian evolution and 
figure \ref{fig1} indeed shows that the situation is different in
this case. $\mu_k(L)$ actually decays as a power law 
with $k$ with an exponent $\alpha_{BSM}=1.30\pm 0.02$ and $n_k(L)$
satisfies the scaling behavior
\beq
n_k(L)=k^{-\beta}f\left(k/L^{1+\zeta}\right)\;\hbox{for
$k>0$} \label{scal}
\eeq
with $\beta_{BSM}=0.58\pm 0.01$.

The scaling function $f(x)$ drops quickly to zero for 
large arguments and tends to a constant $f(0)\cong 
n_1(L)$ for $x\to 0$. The assumption of a single time 
scale $T_\infty(L)$ for a given size is implicit in \req{scal}.
If $\beta<1$, the normalization condition on $n_k(L)$
easily yields the exponent relation
\beq
\zeta=\frac{\beta}{1-\beta}.
\label{zetabeta}
\eeq
This yields $\zeta_{BSM}\simeq 1.44$ in fair agreement with the
direct measure $\zeta_{BSM}= 1.46\pm 0.03$ using eq.\req{TL}.

If $\zeta$ is the indicator of the relevance of memory, the self
organized nature is usually related to the occurrence of 
avalanche events. An avalanche event is made of a spatially and
causally connected series of events. In the BSM the selection
and update of one site at time $t$ may generate RV's that are
smaller than the one that has just been selected. The evolution
will naturally select these RV at time $t+1$. The same may happen
for a certain period and as a result selection events will 
be localized in a small region. 
At each time one avalanche starts so a number of nested avalanches
are active at each time. In the SOC state the duration $s$ of an
avalanche follows a power law distribution $N(s)\sim s^{-\tau}$ that
defines the exponent $\tau$. An avalanche that lasts a time $s$
typically extends on a region $\xi\sim s^h$.

The definition of an avalanche is particularly simple in terms 
of the variables $k_i$. Consider the avalanche started at time 
$t_0$. This will be active at time $t_0+s$ if all sites $i(t)$
selected at times $t_0<t\le t_0+s$ had a counter $k_{i(t)}(t)\le
t-t_0$, i.e. all these sites where generated after the avalanche 
began. On the other hand the selection of a site with a counter
$k$ at time $t$ terminates all avalanches that started after 
time $t-k$. The size of an avalanche that lasts for $s$ time steps
is simply evaluated as the size of the region with counters 
smaller than $k_i\le s$.

The key points we have reached up to now with the introduction 
of counter variables are: {\em i)} dynamics in quenched 
disorder leads to memory effects and is characterized by
power law behavior in both $\mu_k(L)$ and $n_k(L)$. 
{\em ii)} the mechanism of self organization is not the
same in sandpile models\cite{SOC}, that display no memory
effect, and in the BSM. {\em iii)} We can describe both
memory effects and avalanche events in terms of counter 
variables alone. These observations motivates the
introduction of a new model defined in terms of counter
variables alone to study the interplay between 
the effects of memory and self organization.

The model we are going to discuss is defined as follows:
assign a counter variable $k_i$ to each site
$i=1,\ldots,L$ of a $1$ dimensional lattice. At each time
step one site is selected, with a probability $\mu_{k_i}$
that depends on the value of the counter
\beq
\mu_k=\mu_0(k+1)^{-\alpha}.
\label{mukmod}
\eeq 
When a site is selected its counter
variables and that of its neighbor sites are set to
zero. All other variables are increased by one:
\beqar
k_{i+\delta}(t+1)=0\;\;\;\;\;\;\;\;&\;\;&
\hbox{for $\delta=0,\pm 1$}\\
k_j(t+1)=k_j(t)+1&\;\;&\hbox{else}.
\eeqar
The dependence of the selection probability on $k$ is
devised to generalize the situation occurring when
the dynamics is driven by the extreme statistics of a
random field. The larger the time a region has been
tested for selection the smallest the probability it will
be selected. This is the simplest way to account
for a dependence of the local dynamics on the history of
the process and the rules of the model may apply
also to situations where no disorder is present.

For a finite $L$ the system gets to a steady state that is
characterized by a distribution of counters $n_k(L)$ for which
we shall assume the scaling form eq.\req{scal}. Of
course $n_0(L)=3$ since three counters are updated at
each time step. The normalization of the selection
probability $\sum_{k=0}^\infty n_k(L)\mu_k(L)=L$ fixes
the value of $\mu_0(L)$:
\beq
\mu_0(L)=\frac{1}{3+\sum_{k=1}^\infty
n_k(L)(k+1)^{-\alpha}}.\label{mu0}
\eeq

Let us start the discussion of the model from the $\alpha=0$
case. Clearly $\mu_0(L)=1/L$ at all times. The
$n_k(L)$ instead decays exponentially. A simple
explanation of this comes from the 
relation between the number
of sites with $k_i(t+1)=k+1$ and $k_i(t)=k$. 
In the steady state this reads
$n_{k+1}(L)=n_k(L)(1-3/L)$. This immediately yields
$n_k(L)\cong 3\exp(3k/L)$ and $\zeta(\alpha=0)=\beta(\alpha=0)=0$. 
Of course the probability of a connected
event of $s$ steps also goes to zero exponentially with
$s$, i.e. $\tau(\alpha=0)=0$. In conclusion neither memory nor avalanches are
present in the model for $\alpha=0$. 
The same behavior is expected to persist for small values of $\alpha$.
Let us focus on a site with $k_i=k$ and consider the probability
$P_{k}(s)$ that it will not be selected in the next $s$ steps,
under the condition that in this period it will not be updated
because of its neighbors. Clearly $P_k(s)=\prod_{j=k+1}^{k+s}(1-\mu_0
j^{-\alpha})$. It is not difficult to check that, if $\alpha<1$,
$P_k(s)\to 0$ as $s\to\infty$ while it tends to a constant if
$\alpha>1$. So this site, if it is not updated by its neighbors, will
surely be selected sooner or later. This suggests that for $\alpha<1$
the average time $T_\infty(L)$ between two updates of the same
site stays finite and $\zeta(\alpha<1)=\beta(\alpha<1)=0$ (see
eq.\req{zetabeta}). The occurrence of SOC can be excluded as well for
$\alpha<1$. The probability of local, causally connected events
(the avalanches) depends on $\mu_0(L)$. The existence of
such events on all length and time scale requires this
probability to stay finite independent of $L$. Supposing the scaling
form \req{scal} for $n_k(L)$ in eq.\req{mu0}, it is easy to
see that if $\alpha+\beta\ge 1$, as $L\to\infty$,
$\mu_0(L)\to\mu_0(\infty)>0$, while in the opposite case
$\mu_0(\infty)=0$. 

Let us now consider the opposite case: $\alpha=\infty$.
In this case $\mu_k=0\;\forall k>0$ and $\mu_0=1/3$. The model
describes a random walk on a $d=1$ lattice. It is not difficult
to find $\zeta(\infty)=1$ and a distribution $n_k(L)$ that follows 
eq.\ref{scal} with $\beta(\infty)=1/2$. With respect to SOC, the
evolution is a single connected event: every avalanche lasts for an
infinite time. For a finite $\alpha>1$ it is convenient to generalize
the avalanche distribution to account for infinite avalanches:
\beq
N(s)=(1-N_\infty)N_f(s)+N_\infty\delta_{s,\infty}.
\label{Ns}
\eeq
$N_\infty$ is the fraction of avalanches that never stop. The
lack of characteristic lengths in the model suggests that the
distribution of finite avalanches $N_f(s)$ will in general follow
a power law distribution. An
avalanche of duration $s$ is terminated when a site with $k_i>s$ is
selected. If $k_{\rm max}(t)=\max[k_i(t),\,i=1,\ldots,L]$, all the
avalanches that began before time $t-k_{\rm max}(t)$ will never be
terminated. The probability that an avalanche is still active after
$s$ events is 
\beq
P_{\rm act}(s)=\prod_{t=1}^s\left[1-\sum_{k>t}n_k(L)\mu_k(L)\right].
\label{pact}
\eeq
Supposing again eq.\req{scal} for $n_k(L)$, the sum in eq.\req{pact}
goes as $k^{1-\alpha-\beta}$ so that $P_{\rm act}(s)\to 0$ if
$\alpha+\beta(\alpha)<2$. For large $\alpha$ a finite fraction of
avalanche events will never stop while we expect a finite interval
$\alpha\in [1,\alpha_c]$ where $N_\infty(\alpha) =0$ and the usual
scenario of SOC applies. 

\begin{figure}
\vspace{15cm}
\includegraphics{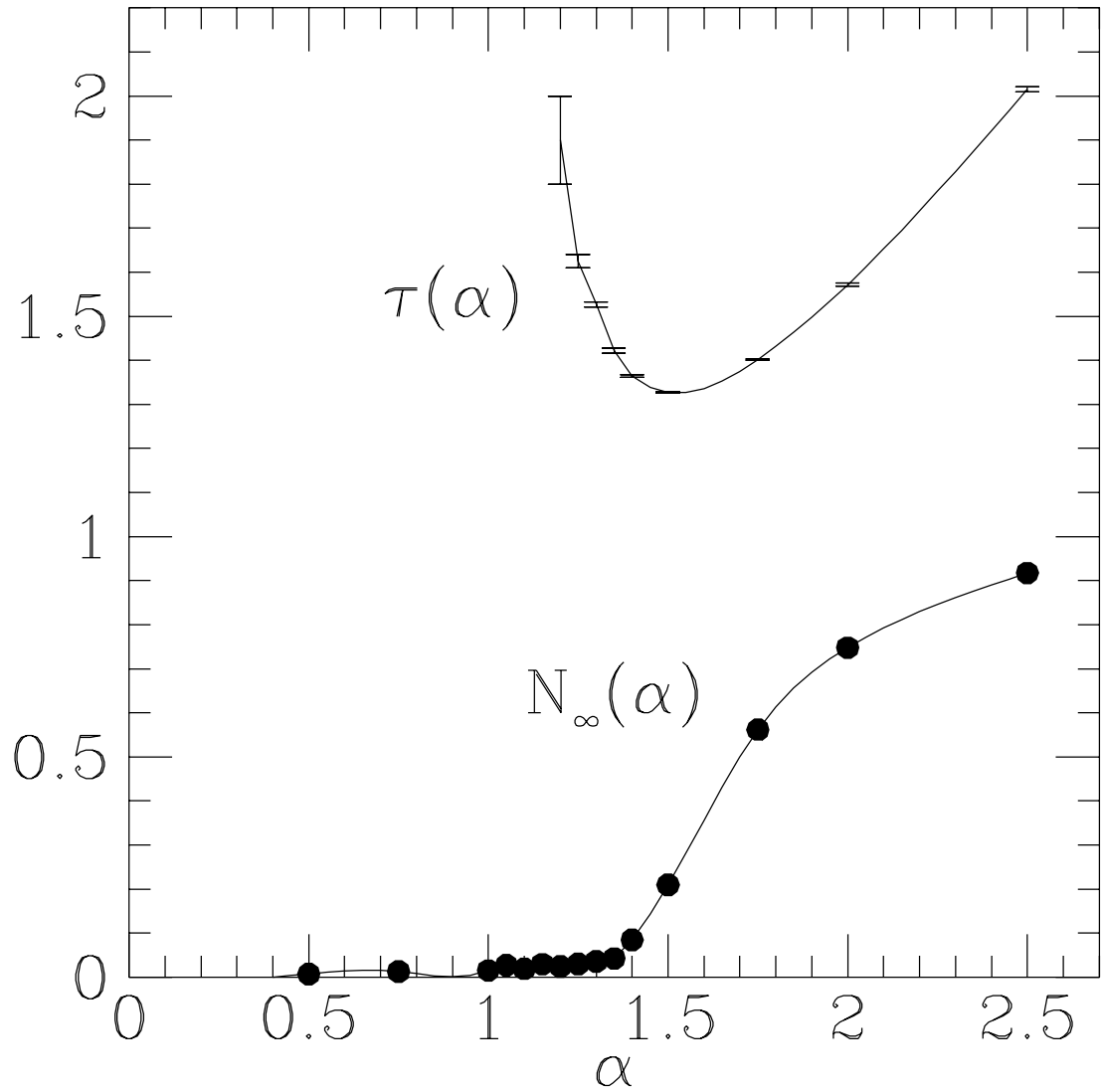}
\caption{Fraction of infinite avalanches $N_\infty(\alpha)$ in a system 
with $L=128$ and $\tau(\alpha)$ exponent.}
\label{fig2}
\end{figure}

This picture was checked by computer simulation of the model.
For $\alpha<1$, as expected, $\mu_0(L)\sim L^{-\omega}$ 
vanishes with $\omega(\alpha)\cong 1-\alpha$.
Table \ref{tab1} lists the values obtained for the exponents
$\beta$ and $\zeta$ by numerical simulations of the
model for sizes up to $L=256$ and $\alpha>1$. The statistical
uncertainty gets large as $\alpha=1$ is approached
from above. The relation \req{zetabeta} is satisfied fairly well.
$\alpha+\beta(\alpha)$ gets bigger than $1$ for $\alpha\cong 
1.4$. Correspondingly, as expected, $N_\infty$ becomes positive.
This is shown in fig.\ref{fig2} where we report also the 
estimate of the exponent $\tau$ obtained for $L=128$ and $256$. 
The distribution $N_f(s)$ was found to decay as a power law for
more than two decades but the value of the exponent actually
decreases as $L$ increases. For small $\alpha$ the
statistics of avalanches gets scarcer and scarcer and a
reliable estimate was not possible. Even though of a qualitative
nature, the behavior of $\tau(\alpha)$ is quite evident from
figure \ref{fig2}. $\tau$ reaches a minimum approximately in the
same region where $\alpha+\beta\cong 1$ and $N_\infty$ starts to
increase. The last occurrence would naturally arises as a result of the
divergence of the normalization integral of $N(s)$ that occurs
for $\tau=1$. This would imply a systematic error in our $\tau$
data of approximately $+0.3$. With this proviso we find that, for
$\alpha=\alpha_{BSM}=1.30$, the exponent should be 
$\tau_{BSM}\approx 1.1$ in good agreement with numerical results
for the BSM\cite{BS,FT}. With respect to the BSM, it is to note
that the values of $\beta$ and $\zeta$ agree with those
obtained for $\alpha=1.30$. We confirmed numerically that
$N_\infty(L)\sim L^{-0.98}$ goes to zero for the BSM.

\begin{table}
\begin{tabular}{ccc} 
{$\alpha$} &{$\beta$} & $\zeta$ \\ 
\tableline
1.10 & 0.3   $\pm$ 0.1   & 0.5	 $\pm$ 0.2  \\
1.20 &	0.48  $\pm$ 0.01  & 0.90 $\pm$ 0.04 \\
1.30 &	0.58  $\pm$ 0.01  & 1.40 $\pm$ 0.06 \\
1.40 &	0.619 $\pm$ 0.005 & 1.53 $\pm$ 0.03 \\
1.50 &	0.613 $\pm$ 0.005 & 1.47 $\pm$ 0.03 \\
1.75 &	0.571 $\pm$ 0.005 & 1.31 $\pm$ 0.03 \\
2.00 &	0.545 $\pm$ 0.005 & 1.17 $\pm$ 0.02 \\
2.50 &	0.510 $\pm$ 0.005 & 1.06 $\pm$ 0.02 \\
\end{tabular}
\label{tab1}
\end{table}


\begin{thebibliography}{99}

\bibitem{SOC} P. Bak, C. Tang and K. Weisenfeld, Phys.
Rev. Lett. {\bf 59}, 381, (1987); Phys. Rev. A {\bf 38},
364 (1988).

\bibitem{kardar} T. Hwa, M. Kardar: Phys. Rev. Lett. {\bf 62}, 1813 (1989)

\bibitem{LPSOC} L. Pietronero, A. Vespignani and S. Zapperi,
Phys. Rev. Lett. {\bf 72}, 1690 (1994).

\bibitem{BSRG} M. Marsili: {\em Renormalization Group approach to the self 
organization of a simple model of biological evolution} submitted to
 Europhys. Lett. 1994.

\bibitem{FT} M. Paczuski, S. Maslov and P. Bak {\em Field Theory
of Self Organized Criticality} preprint.

\bibitem{earthq} J. M. Carlson and J. S. Langer: Phys. Rev. Lett. {\bf 62},
 2632 (1989)

\bibitem{aging} M. Lederman, R. Orbach, J. M. Hamman, M. Ocio and
E. Vincent: Phys. Rev. B {\bf 44}, 7403 (1991) and references
therein.

\bibitem{bark} J. P. Sethna, K. Dahmen, S. Kartha, J. A. Krumhansl, B. W.
Roberts and J. D. Shore Phys. Rev. Lett. {\bf 70}, 3347
(1993).

\bibitem{snepp} K. Sneppen: Phys. Rev. Lett. {\bf 69}, 3539
(1992), {\bf 71}, 101 (1993).

\bibitem{interf} L.-H. Tang and H. Leshhorn: Phys.
Rev. A {\bf 45}, R8309 (1992), S. V. Buldyrev, A.-L.
Barab\'asi, F. Caserta, S. Havlin, H. E. Stanley and
T.Vicsek: Phys. Rev. A { \bf 45}, R8313 (1992).

\bibitem{BS} P. Bak and K. Sneppen: Phys. Rev. Lett. {\bf
71}, 4083 (1993), H. Flyvbjerg, K. Sneppen and P. Bak:
Phys. Rev. Lett. {\bf 71}, 4087 (1993).

\bibitem{IP} R. Lenormand and S. Bories: C. R. Acad. Sci.
{\bf 291}, 279 (1980), R. Chandler, J. Koplick, K. Lerman,
J. F. Willemsen: J.Fluid Mech. {\bf 119}, 249 (1982), D.
Wilkinson and J. F. Willemsen: J. Phys. A {\bf 16}, 3365
(1983).

\bibitem{cdw} G. Kriza and G. Mih\'ali: Phys. Rev. Lett. {\bf
56}, 2529 (1986); R. M. Fleming and L. F. Schneemeyer: Phys. Rev. B
{\bf 33}, 2930 (1986); G. Gruner: Rev. Mod. Phys. {\bf 60}, 1129 
(1988).

\bibitem{RFIM} R. Bruinsma and G. Aeppli: Phys. Rev. Lett. {\bf
52}, 1547 (1984); J.Koplick and H. Levine: Phys. Rev. B {\bf 32},
280 (1985); B. Koiller, H. Ji and M. O. Robbins: Phys. Rev. B
{\bf 46}, 5258 (1992).

\bibitem{abel} D. Dhar : Phys Rev. Lett.  {\bf 64}, 1613 (1990)

\bibitem{RTS} M. Marsili: {\em Run Time Statistics in
Models of Growth in Disordered Media}. To appear in Nov.
1994 issue of J.Stat.Phys.

\bibitem{metro} N. Metropolis, A. W. Rosenbluth, M. N.Rosenbluth, A. 
M. Teller, and E. Teller: J. Chem. Phys. {\bf 21}, 1087 (1953). 

\end{thebibliography}
\end{document}